\documentclass[a4paper,12pt]{article}
\usepackage{graphicx}
\begin{document}
\vspace*{-0.7cm}

\begin{center}

   {\large\bf  Quantum mechanics and irreversible time flow
\footnote {Presented at the conference  "Irreversible quantum
dynamics", ICTP Trieste, 29.7.-2.8.2002} } \\[1mm]
        Milo\v{s} V. Lokaj\'{\i}\v{c}ek \\
     Institute of Physics, AS CR, 18221  Prague 8, Czech Republic \\
\end{center}

{\bf Abstract }

 Time flow has been embodied in time-dependent Schr\"{o}dinger
equation representing one of the foundations of quantum mechanics.
Pauli's criticism (1933) has, however, indicated that the
assumptions concerning representation Hilbert space have led to
some contradictions. Many authors have tried to solve this
discrepancy practically without any actual success. The reason may
be seen in that two different problems have been mixed: the
problem of Pauli (being more general and more important) and
non-unitarity of exponential phase operator of linear harmonic
oscillator introduced by Dirac, as demonstrated in 1964. The
problem will be discussed in a broad historical context and a
solution based on extension of representation Hilbert space will
be shown.   \\

 { \bf  1. Introduction  }

Time flow in quantum mechanics represented an important problem
practically during the whole past century. It is, therefore,
necessary to start with a brief survey concerning the beginning of
quantum mechanics. This beginning is linked closely with the name
of E. Schr\"{o}dinger (1925, \cite{schr}), who believed firmly in
wave nature of matter. He proposed his famous wave equation and
was successful when he was able to reproduce all main results
obtained earlier with the help of Hamilton equations. It was
possible to derive all characteristics of a physical system from a
wave function $\Psi({\tilde x},t)$ where ${\tilde x}$ represented
coordinates of all matter objects. A great success has been seen
in that the discrete atom energy levels have corresponded to
eigenvalues of Hamiltonian \cite{schr2}. The wave function $\Psi$
was interpreted as probability distribution by M. Born
\cite{born}. N. Bohr  \cite{bohr28} attributed then the
probabilistic properties directly to individual matter particles;
Heisenberg's \cite{hei} uncertainty relations being linked closely
with their properties.

 A further step was done by J. von Neumann \cite{vonne} who showed that
individual $\Psi$-functions at given $t$ values may be represented
by vectors in the Hilbert space spanned on eigenfunctions of the
corresponding Hamiltonian (determining the total energy of a given
physical system). Some additional assumptions have been
introduced, however, into the corresponding mathematical model.
And Pauli's criticism \cite{pauli} has concerned just them. Pauli
has showed that under the mentioned conditions the time evolution
(introduction of the time operator) has required for the
Hamiltonian to possess continuous spectrum belonging to the whole
real interval $({-\infty,+\infty})$, which contradicts the
necessity of energy being positive (or at least limited from
below).

This problem has been, however, often related to another one
concerning the non-unitarity of exponential phase operator
  \begin{equation} \label{exph}
 \mathcal{E} \;\;= \;\; e^{-i\Phi}\;\;   %% =\;\; e^{-i\omega T}
\end{equation}
where the phase $\Phi$ is proportional to time. The wxponential
phase operator for linear harmonic oscillator was defined by Dirac
\cite{dirac} already in 1927 and it was assumed to be unitary.
However, it was shown later by Suskind and Glogower \cite{sus}
that the operator $\mathcal{E}$ defined in such a way was
isometric only; see also the review by Lynch \cite{lynch}. The
non-unitarity has been then assumed to follow from the same source
as Pauli's criticism. It will be shown, however, in the following
that they must be handled as two separate problems.

{\small Both the problems will be described to a greater detail in
Sec. 2. Pauli's problem is closely related to two basic
assumptions of quantum-mechanical model: time-dependent
Schr\"{o}dinger equation and superposition principle (as will be
explained in Sec. 3), while the other problem is more formal and
relates rather to some features of quantum field theory. And we
should ask how it was possible that the standard
quantum-mechanical model with known paradoxes was regarded as the
only model of microworld during the whole past century, even if
critical arguments have been repeatedly formulated (e.g.,
Einstein's criticism \cite{ein}); the corresponding story being
described in Sec. 4. In Sec. 5 it will be then shown that it has
not been only the known mistake of von Neumann, but also two other
mistakes that have played yet more important role in common
conviction and in commonly accepted conclusions.

 The possible solutions of both the problems (concerning the
introduction of time operator and avoiding the mistakes described
in Sec. 5) will be given in the other half of the paper. In Sec. 6
it will be shown that Pauli's problem may be solved by extending
the Hilbert space according to the theory of Lax and Phillips. The
non-unitarity of exponential phase operator may be then solved by
doubling further such (already once extended) Hilbert space
according to the proposal of Fajn, or of Newton and Bauer, as
shown in Sec 7. Some differences in the evolution of states
belonging to discrete or continuous spectra of Hamiltonian will be
discussed in Sec. 8. Several concluding and summarizing remarks
will be given in Sec. 9.  }  \\

  {\bf  2. Quantum mechanics and time operator  }

As already mentioned the problem of time operator in quantum
mechanics started to be solved by Dirac \cite{dirac} for the case
of linear harmonic oscillator described with the help of
Hamiltonian
\begin{equation}
  H \;=\; \frac{p^2}{2m} \;+\; kq^2 \;            \label{lioh}
\end{equation}
when annihilation and creation operators
\begin{equation}
    a \;=\; p - im\omega q, \;\;\; a^{\dag} \;=\;  p+im\omega q
\end{equation}
have fulfilled relations
 \begin{equation}
    [H,a] \;=\; -\omega a\:, \;\;\; [H, a^\dag] \;=\; \omega  a^{\dag}\:;
                    \;\;\;\; \omega=\sqrt{\frac{k}{m}} \; .
\end{equation}
 It has been possible to introduce operators
\begin{equation}
 \mathcal{E}\;=\; (a a^\dag+1)^{1/2}a\:,
             \;\;\;  \mathcal{E}^\dag  \;=\; a^\dag (aa^\dag +1)^{1/2})
\end{equation}
fulfilling the relations
 \begin{equation}
 [H,\mathcal{E}] \;=\; -\omega \mathcal{E}\;, \;\;\;
          [H,\mathcal{E}^\dag] \;=\; +\omega \mathcal{E}^\dag \;
 \end{equation}
 representing exponential phase operator corresponding to Eq.(\ref{exph}).

However, it has been shown later by Susskind and Glogower
\cite{sus}  that the operator  $\mathcal{E}$ is not unitary, but
only isometric, as it holds
 $\mathcal{E}^\dag\mathcal{E}\:u_{1/2} \;\equiv\;  0$
(and not $u_{1/2}$). It means that the unitarity condition is not
fulfilled for the state vector corresponding to the minimum-energy
(vacuum) state. And we may ask whether the problem may be
correlated to the problem of Pauli or not.

As to Pauli's criticism it represents more important problem.
Starting from the standard quantum-mechanical model the time
evolution of a physical system is fully determined by the wave
function
\begin{equation}
 \Psi({\tilde x},t) \;=\; \int dE\: a_E(t)\:u_E({\tilde x})\:    \\
\end{equation}
where $u_E({\tilde x})$ are eigenfunctions of corresponding
Hamiltonian. And according to Pauli the existence of time operator
in the corresponding Hilbert space requires for Hamiltonian to
have continuous spectrum $E\;\in\;(-\infty,+\infty)$.

And one can hardly regard these two discrepancies as consequences
of one common problem. Each of them must be removed in a proper
way. However, before being possible to propose the  solutions of
these two problems it is necessary to go back to foundations of
quantum mechanics and to the corresponding story in the 20th
century.
\\

 { \bf  3.  Time-dependent Schr\"{o}dinger equation and superposition principle  }

It is possible to say that the standard quantum mechanics is
based on two following basic assumptions:  \\
 \hspace*{2mm}  - any state of a physical system and its time
evolution is represented by the wave function
$\psi(\{x_{k,j}\},t)$ that is obtained by the solution of
time-dependent Schr\"{o}dinger equation
\begin{equation}
   ih\frac{\partial}{\partial t}\psi(\{x_{k,j}\},t) \;=\; H\:\psi(\{x_{k,j}\},t)
                                                     \label{scht}
\end{equation}
where  $H$ is corresponding Hamiltonian
 \begin{equation}
 H \;=\; \Sigma^N_{j=1}\Sigma_{k=1}^3\frac{p^2_{k,j}}{2 m_j} \;+\;
                                 V(\{x_{k,j}\})       \label{ham}
\end{equation}
and $\{x_{k,j}\}$ and $\{p_{k,j}\}$ are coordinates and momentum
components of individual objects; $V(\{x_{k,j}\})$ is the sum of
potential energies between $N$ individual mass objects;  it is put
$\{p_{k,j}\} \;=\; \frac{\hbar}{i}\frac{\partial}{\partial
x_{k,j}} \;$ in Eq. (\ref{scht}) ;  \\
 \hspace*{2mm}  - any physical state is represented by a vector
in Hilbert space being spanned on one set of Hamiltonian
eigenfunctions
 \begin{equation}
        H\: \psi_E(\{x_{k,j}\}) \;=\; E\: \psi_E(\{x_{k,j}\})
 \end{equation}
and all states are bound together with the help of superposition
principle.

However, these two assumptions exhibit mutual contradiction, if
applied to a physical system. It has been argued that any
superposition of two solutions of Schr\"{o}dinger equation has
been again a solution of the same equation. However, such a
statement is entitled and has regular physical meaning only if
both these solutions correspond to the same initial conditions.
Superposing solutions belonging to different initial conditions
one obtains solutions corresponding to fully different initial
conditions, which means that significantly different physical
states have been combined in a unallowed way.

 The general mathematical superposition principle holding for
linear differential equations has nothing to do with physical
reality, as actual physical states and their evolution is uniquely
defined by corresponding initial conditions. These initial
conditions characterize individual solutions of Schroedinger
equation (\ref{scht}); they correspond to different properties of
a physical system, some of them being conserved during the whole
evolution.

 The physical superposition principle has been deduced from
the linearity of Schr\"{o}dinger differential equation without any
actual reason.  This drastic assumption was introduced into the
physics without any need and any proof; the solutions belonging to
diametrically different initial conditions have been arbitrarily
superposed. Statements that quantum mechanics (including
superposition rules) has been experimentally verified must be
regarded as wrong. All hitherto tests have concerned consequences
following from Schr\"{o}dinger equations only.

However, the story of quantum mechanics in the 20th century was
paved by other more serious mistakes. One of them (i.e., the
mistake of von Neumann) has been usually well known, while the
other two have remained practically hidden till now. We shall
follow the whole story in the next section.  \\

 { \bf  4. Quantum mechanics in the 20th century  }

 The foundations of quantum mechanics have been mentioned already
in Sec. 1. They have related mainly to the names of E.
Schr\"{o}dinger (1925), M. Born (1926), W. Heisenberg (1927), N.
Bohr (1927) and J. von Neumann (1932). A deep ontological change
has been introduced into matter nature without any actual physical
reason. Paradoxical ideas have been accepted by the most
physicists even if A. Einstein, who discovered the photon
\cite{ein2} and should be held for one of main founders of quantum
theory, criticized strongly physical characteristics following
from the given quantum-mechanical mathematical model. Critical
comments ("circle" proof) have been published also by G. Herrmann
\cite{herr}, but they have not been taken into account, either.

 Einstein \cite{ein} denoted the given model as incomplete and
required to add some other parameters, the so called hidden
variables, necessary for a full description of concerned physical
systems. His critical arguments have been, however, refused by N.
Bohr \cite{bohr} and practically by the whole physical community,
as well.

Nevertheless, they have been permanently discussed during the
whole century. The discussion started to continue when D. Bohm
\cite{bohm} showed in 1952 that a kind of hidden variable was
contained already in the time-dependent Schr\"{o}dinger equation.
A decisive progress came then when J. Bell \cite{bell} specified
exactly and modified non-physical assumption (made use of by von
Neumann) and derived his known inequalities. It was believed that
the time came when the controversy between Einstein and Bohr might
be solved on experimental basis. And really some suitable
experiments (measurement of coincidence transmission of polarized
photons through a pair of polarizers) have been proposed and also
performed. The final results provided by L. Aspect et al.
\cite{asp} in 1982 came then to the following conclusions:  \\
 \hspace*{2mm} - Bell's inequalities have been violated by
experimental data;  \\
 \hspace*{2mm} -  the experimental results have been practically
in agreement with quantum-mechanical predictions.

It might seem that the given problem has been solved by these
experiments. In the first years it was actually spoken about the
victory of quantum mechanics. However, the basic logical
controversy has remained unsolved and the discussions have
continued until now.

While the mistaking assumption of von Neumann has concerned
general mathematical structure (measurement problem in quantum
mechanics) all other discussions has concerned concrete points of
experimental arrangement: coincidence transmission of equally
polarized photons through two polarizers. And one must ask whether
all arguments (or assumptions) made use of in interpretation of
these coincidence measurements have corresponded to reality.

The two additional arguments, which may be considered misleading
in a similar way as that of von Neumann, seem to have played a
very decisive role in the common acceptance of given conclusions.
One (not reasoned) assumption has been introduced in deriving
Bell's inequalities. And a seemingly convincing (but false)
argument was given in the book by Belifante in 1973 \cite{belif}.
Both these arguments will be analyzed in the next section.   \\

 { \bf 5. Mistaking arguments  }

It is possible to say that the way to paradoxical interpretation
of quantum mechanics has been fundamentally influenced by tree
mistakes. The consequences of von Neumann's mistake have lasted
practically until now, even if it has been recognized and removed
in principle already earlier. It has had a close relation to the
disagreement between the physical superposition principle and the
time-dependent Schr\"{o}dinger equation. Not only the wave
function but also its first derivatives must be known at one time
moment if the evolution of a physical system should be determined.

It means that all properties and the whole evolution cannot be
fully defined by one vector of the standard Hilbert space if
something more is not added from outside, e.g. superposition
coefficients defined as functions of coordinates taken from
corresponding solutions of Schroedinger equation. A momentum
direction cannot be derived from a mere linear combination of
eigenvectors if energy corresponds to momentum squared. All
properties may be determined by one vector only if the Hilbert
space is correspondingly doubled (see Sec. 6).

However, in the last time the public conviction has been
influenced more by two other arguments that must be denoted as
misleading. The mistake of von Neumann has been removed by J. Bell
only partially. A similar (even if much weaker) assumption has
been involved in derivation of Bell's inequalities, as well. These
inequalities were being derived in different ways (see, e.g.,
\cite{au2}), but in all an important (latent) assumption has been
involved.

We will comment now very briefly two kinds of their derivations
(shown, e.g., in Ref. \cite{au2}), which have been discussed to a
necessary detail in Ref. \cite{loka}. In the first approach it is
possible to pass from Eq. (3.11) to Eq. (3.12) (in Ref.
\cite{au2}) only if the probabilities (or expectation values) in
different pair measurements might be interchanged, which means
that the influence of internal structures of both the polarizers
(e.g., the influence of the impact of individual localized photons
into the atom grid) has been neglected. Bell's inequalities cannot
be derived if this degree of freedom is taken into account; in
addition to photon polarization and polarizer axis orientation
(i.e., if all three vector characteristics are respected). In the
other approach (discussed in \cite{loka}, as well) practically the
same neglection has been involved in the passage from Eq. (3.14)
(or (3.15)) to Eq. (3.16) (see Ref. \cite{au2}).

It means that the measuring device has been regarded as a black
(or at least semi-black) box in a similar way as in the standard
quantum mechanics. Actual localization of photon impact into
measuring device has been fully omitted. And one must conclude
that the violation of Bell inequalities in coincidence
polarization experiments is quite irrelevant as to the support for
standard quantum mechanics.

The quantum-mechanical interpretation of EPR experiments seems,
however, to have been much more supported by the argument of
Belifante who stated that the predictions of a hidden-variable
theory had to differ significantly from those of
quantum-mechanical model (see the corresponding graph on p. 284 in
Ref. \cite{belif}), which is not true. According to a
hidden-variable theory the transmission of a photon (or of two
equally polarized photons in coincidence arrangement) through a
polarizer pair should equal
\begin{equation}
 m(\alpha)\;=\; \int_{-\pi/2}^{\pi/2}p_1(\lambda)
         \;p_1(\lambda-\alpha)\;d\lambda            \label{mpp}
\end{equation}
where $p_1(\lambda)$ is transmission probability through one
polarizer; $\lambda$ - deviation of photon polarization from the
axis of the first polarizer; and $\alpha$ - the angle deviation of
the second polarizer. The same formula (\ref{mpp}) holds exactly
in both the arrangements if the photon polarization does not
change in passing through a polarizer.

It is known from the one-side arrangement that it holds
\begin{equation}
 m(\alpha)\;=\; (1-\varepsilon)\cos^2(\alpha)+\varepsilon  \label{gmal}
\end{equation}
where for real (imperfect) polarizers is $\varepsilon > 0$
(generalized Malus law); $\varepsilon$ being very small. Belifante
has chosen
\begin{equation}
    p_1(\lambda) \;=\; \cos^2(\lambda);
\end{equation}
i.e., he has interchanged quite arbitrarily the transmission
through one polarizer and through a pair of them. Malus law may be
reproduced easily also in hidden-variable alternative  if
$p_1(\lambda)$ is chosen  in a corresponding way as shown in Fig.
1 (full line).

\begin{figure}[htb]
\begin{center}
\includegraphics*[scale=.4, angle= -90]{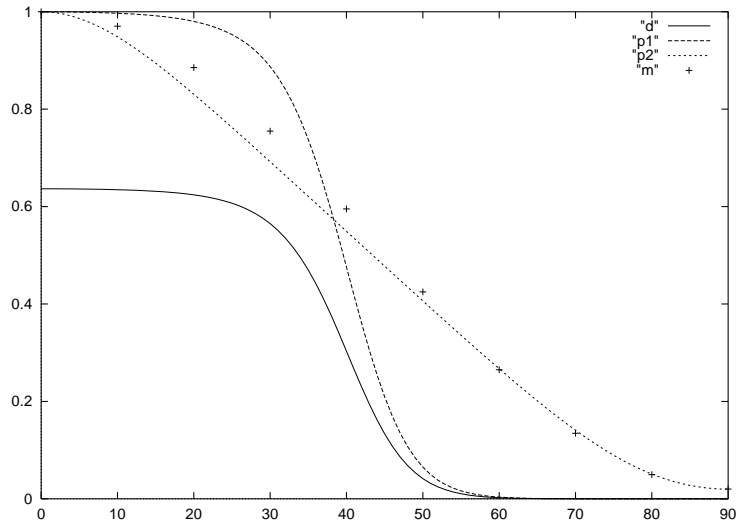}
   \caption{ {\it
Transmission probability through a polarizer pair leading to Malus
law;   $p_1(\lambda)$ - full line;  $m(\lambda)$ - dashed line;
             Malus law - individual points.   } } \label{fg1}
\end{center}
 \end{figure}
\vspace{0.2cm}

And one must conclude that any preference for the
quantum-mechanical interpretation of coincidence EPR experiments
does not follow from experimental data. Nothing prevents us from
interpreting all available experiments on the basis of a
hidden-variable theory. This fact has opened also a new way for
answering the question of the time operator and/or of Pauli's
problem.       \\

 {\bf  6.  Pauli's problem and theory of Lax and Phillips }

In the standard quantum-mechanical model the Hilbert space has
been spanned on a simple vector basis consisting of Hamiltonian
eigenfunctions (independently of spectrum kind). In such a case
the states corresponding to momenta of different signs are
represented in principle by the same vector (Hamiltonian being a
function of momentum squares). The sign of wave function
derivatives depends on the choice of coordinate system and must be
always added from outside; additional information taken from
Schroedinger equation. And Pauli's critique \cite{pauli} has
concerned just this fact. The given deficiency may be removed if
the standard Hilbert space is doubled in a way, as it was done by
Lax and Phillips \cite{lax1} already in 1967 (see also
\cite{lax2}) and derived independently by Alda et. al. \cite{alda}
in solving the problem of unstable particles exhibiting purely
exponential decay.

Let us demonstrate at least shortly the given Hilbert structure on
the example of a system consisting of two free particles. The
corresponding Hilbert space consists then of two subspaces:
\begin{equation}
      \mathcal{H}  \;\equiv\;  \{\Delta^-,\Delta^+\}
\end{equation}
that are mutually related by evolution operator
\begin{equation}
            U(t) \;=\; e^{-iHt} \;\;  (t \ge 0).
\end{equation}
 It holds, e.g.,
\begin{equation}
\mathcal{H} \;=\; \overline{\Sigma_t U(t)\Delta^-}
                             \;=\;\overline{\Sigma_t U(-t)\Delta^+}
\end{equation}
Individual subspaces $\Delta^-$ and $\Delta^+$ are spanned on
Hamiltonian eigenfunction in the usual way. Any $t$-dependent wave
function obtained by solving Schr\"{o}dinger equation may be then
represented by a trajectory corresponding to given initial
conditions.

In the case of continuous Hamiltonian spectrum (free particles)
any point on such a trajectory may be characterized by expectation
values of the operator $R\;=\;\frac{1}{2}\{{\bf p}.{\bf q}\}$,
where ${\bf q}$ and ${\bf p}$  are coordinates and momentum
components of one particle in CMS. The states belonging to
$\Delta^-$ are incoming states, and those of $\Delta^+$ - outgoing
states (independently of the chosen coordinate system).
 The evolution goes always in one direction from "in" to "out".

As these two different kinds of states may be experimentally
distinguished  it is useful to separate "in" and "out" states into
two mutually orthogonal subspaces. It is then also possible to
join an additional orthogonal subspace that might represent
corresponding resonances formed in particle collisions (see
\cite{alda}); i.e.
\begin{equation}
 \mathcal{H} \;\equiv\; \{\Delta^-\oplus\Theta\oplus\Delta^+\}\;.
\end{equation}
It is only necessary to define the action of evolution operator
between $\Theta$ and other subspaces in agreement with evolution
defined already in individual $\Delta^\pm$.

The evolution goes  in one direction, at least from global view;
some transitions between internal states of $\Theta$ may be
reversible and chaotic. However, global trajectories tend always
in one direction;  see the scheme in Fig. 2.
 \\[0.5cm]

  \hspace*{9.0cm}   { $R \;=\; \frac{1}{2}\{\mathbf{p}.\mathbf{q}\};   \;\;
                                     \langle i[H,R]\rangle > 0$ }  \\
 \hspace*{3.5cm}  $\Delta^{(-)}\; \hspace{1.7cm} |
                                            \hspace{0.8cm} \;\Delta^{(+)}$ \\
 \hspace*{3.2cm}  { $\langle R\rangle <0 \hspace{1.4cm} |
                                      \hspace{0.8cm}\langle R\rangle >0$  \\
 \hspace*{3.1cm} { "in" } \hspace{0.1cm} $\searrow
                                   \hspace{0.4cm}     \longrightarrow  |
                               \hspace{0.8cm} { "out" } \longrightarrow$ \\
 \hspace*{3.0cm}  $ \_\_\_\_\_\_\_ \_\_\_\_\_\_\_\_\_\_\_\_\_\_|
            \_\_\_\_\_\_\_\_\_\_\_\_\_\_\_\_\_\_ $ \\
   \hspace*{4.5cm}$| \hspace{1.6cm}  \nearrow \hspace{1.1cm}  | $    \\
 \hspace*{4.5cm}$| \hspace{1.5cm} { \Theta}\hspace{1.6cm}| $      \\
 \hspace*{4.5cm}$|\_\_\_\_\_\_\_\_\_\_\_\_\_\_\_\_\_\_\_\_\_\_\_|    $ }      \\[1mm]
 Fig.2: {\it Scheme of the Hilbert space (for a two-particle system)
extended according to Lax and Phillips; three mutually orthogonal
subspaces and direction of time evolution (continuous specrum). }
\\

In the case of discrete Hamiltonian spectrum (e.g., harmonic
oscillator) the wave function has similar t-dependent form.
However, the evolution is periodical as a rule. The Hilbert space
will consist now of two non-orthogonal subspaces (or of an
infinite series of such pairs if one wants to number individual
periods). The evolution may be again characterized by trajectories
corresponding to different initial conditions. Different points of
these trajectories may be now determined by expectation values of
the phase operator from the interval $(0,2\pi)$ in individual
(neighbour) subspace pairs; or in analogy with the continuous case
(for one period) by those of $tg(\Phi/2)$ or $cotg(\Phi/2)$
increasing again from $-\infty$ to $+\infty$.

Such an extended model enables to represent the time evolution in
the Hilbert space in a full agreement with actual behavior of
physical systems. It gives the full answer to Pauli's criticism
that was misunderstood in the past having been related to
non-unitarity of exponential phase operator found
in the case of linear harmonic oscillator.  \\

  { \bf  7.   Exponential phase operator and its unitarity}

We have already mentioned  that the exponential phase operator
defined by Dirac in the case of linear harmonic oscillator is not
unitary. Unitarity breaks at minimum-energy (vacuum) state, which
means that the phase operator is not regularly defined in the
whole standard Hilbert space. The problem  concerns the
Hamiltonian possessing discrete spectrum. And the question may be
put, whether the problem exists also in the case of continuous
spectrum when the vacuum state corresponds practically to zero
energy.

As to  the discrete spectrum the response was given by Fain
\cite{fajn} already in 1967, who showed that the problem may be
solved by doubling the standard Hilbert space. The doubled Hilbert
space should consist of two identical mutually orthogonal
subspaces:
\begin{equation}
   \mathcal{H} \;=\; \mathcal{H}_+ \oplus  \mathcal{H}_- \;   ,
\end{equation}
the orthogonality of them remaining conserved during the whole
time evolution:
\begin{equation}
 U(t)\mathcal{H}_+ \in \mathcal{H}_+,\;\; U(t)\mathcal{H}_- \in \mathcal{H}_- .
\end{equation}
 The minimum-energy states in both subspaces
have been mutually linked with the help of the annihilation-type
operators. The states in the two mutually orthogonal subspaces
(with separated time evolution) may be distinguished only by
different signs in the relation between the phase and the flowing
time: $\Phi \;=\;\pm \omega T$.

Newton \cite{newt} and Bauer \cite{bauer} have proposed a similar
solution independently later. However, they have not recognized
the role of phase sign; he regarded the other subspace as a ghost
space without any physical meaning.

In fact the given sign may be interpreted probably in a physical
sense. It follows from the analysis of three-dimensional
oscillator (see \cite{pk}) that it may be related to the
orientation of the co-ordinate system or to the orientation of the
corresponding component of resulting spin (or of angular momentum
of a two-particle system). The individual Hilbert subspaces
$\mathcal{H}_+$ and $\mathcal{H}_-$ should be, of course, extended
according to the proposal of Lax and Phillips (see Sec. 6).  \\

  { \bf  8. Twice extended Hilbert space }

Both the problems concerning regular description of time evolution
in the framework of Hilbert space could be solved as two different
problems by suitable extensions of the standard Hilbert space. The
solution could be hardly reached in the past when the problems
were combined together.

The twice extended  model may provide a regular mathematical basis
for the description of irreversible processes in the microscopic
physics. All objects are described semiclassically, which enables
to take into account also the microscopic characteristics of
measuring devices, as needed.

The extension proposed by Lax and Phillips should be regarded as
basic and more important. It allows to represent truly
corresponding solutions of time-dependent Schr\"{o}dinger equation
for both the kinds of spectrum (continuous and discrete). The
physical evolution operator $U(t)$ moves vectors in one direction
only; from "in" to "out" (evolution into infinitum) in the
continuous case, and alternatively from one subspace to another in
the discrete periodical case. Points on individual evolution
trajectories may be characterized by expectation values of
operators $R$ or $\Phi$ (resp. $tg(\Phi/2)$ in one half-period).
The given characteristics for the discrete spectrum has been
derived by solving the problem of three-dimensional harmonic
oscillator \cite{pk} since the simplified picture in the linear
case could not provide a sufficient answer.

As to the continuous case it is useful to introduce the
orthogonality condition between "in" and "out" subspaces as
considered by Lax and Phillips. An additional subspace may be
included that may represent corresponding unstable resonances
formed in collision processes. The problem will be mentioned yet
in the next section (conclusion).

To remove the non-unitarity of exponential phase operator a
further doubling of the Hilbert structure in addition to that
required by Pauli's problem is necessary. This second doubling
differs from the preceding approach in that the two subspaces must
be now permanently orthogonal; there is not any linkage of theirs
by evolution operator, either. The evolution trajectories do not
leave individual orthogonal subspaces. And one should ask whether
these two independent evolutions represent identical or different
physical processes. One can hardly expect any measurable
difference (and distinguishing) in a (two-particle) system
corresponding to a continuous spectrum. However, a question has
remained open in the case of discrete spectrum as the sign of the
phase may relate closely to the space orientation of angular
momentum.    \\

  { \bf  9. Conclusion  }  \\

It has been shown that the paradoxical quantum-mechanical picture
accepted by the physical community in the past century was
supported by several mistakes in corresponding argumentations. A
different picture of microworld following from all experimental
data if these mistakes have been removed should be now accepted;
being in agreement with Einstein's view. It is possible to
represent corresponding evolution trajectories in a Hilbert space
if this space is extended (in principle doubled) in comparison to
that used standardly in quantum mechanics. There have been, of
course, two different problems that have had to be solved
separately.

As to the Pauli's problem it could be fully solved by introducing
the extension proposed by Lax and Phillips in 1967. It has been
then possible to represent physical evolution by trajectories
characterized by different initial conditions. However, a kind of
superselection rules should be applied to states belonging to
trajectories corresponding at least to different values of
physical quantities being conserved during the evolution. Physical
systems (consisting of stable objects) exhibit the evolution that
may be derived from Schr\"{o}dinger or Hamilton equations.
 The points on individual evolution trajectories may be then
characterized by expectation values of operators $R$ or $\Phi$
being in unique relation to the time operator. It means that
evolution trajectories exhibit similar properties as those in
classical phase space.

The other problem concerning the non-unitarity of exponential
phase operator should be regarded as less important. It was solved
by Fain also in 1967, at least in principle. However, the solution
was demonstrated with the help of linear harmonic oscillator where
its full physical meaning has remained hidden. It may be
understood only when the approach is applied to a
three-dimensional system (harmonic oscillator). The question
concerns the meaning of the phase sign (relatively to flowing
time). It has been shown that it may have a sense for bound
systems (with discrete Hamiltonian spectrum) as it may be related
to the orientation of angular momentum. However, a final question
of actual physical meaning has remained yet open and should be
further analyzed.

And it is possible to conclude:

(i) The evolution of all matter world may be described as
continuous and irreversible.

(ii) The existence of quantum energy states cannot be regarded as
a support for quantum jumps (denoted as damned by
Schr\"{o}dinger); however, they do not represent any support for
waves, either.

 It is evident that all evolution goes always in an irreversible
way, i.e. from its origin to its end. It concerns living as well
as non-living objects. Only exactly periodical systems may return
to some preceding states, which holds, of course, when these
systems are mutually isolated; anyway, there is not any reversible
evolution. When they are a part of greater systems the total
evolution should be always described as fully irreversible. And
the mathematical model based on the given extension is able to
involve the description of complex physical systems in agreement
with reality.

The given approach seems to be in agreement also with the recent
proposal of W. Lamb \cite{lamb} who has tried to interpret the
quantum-mechanical approach in a semiclassical way. The only
difference seems to consist in that some serious problems have
been removed by narrowing the validity of superposition principle
(by further extension of validity of superselection rules).
 The new interpretation of quantum-mechanical approach is also
decisively supported by experimental results gained in measuring
light transmission through three polarizers (see \cite{loka2}).
\\

{\footnotesize

\end{document}